\documentstyle[12pt,dir_apj/aasms4,epsf]{article}
\def\sun{_\odot}

\def\Dwa{$\,$\uppercase\expandafter{\romannumeral5}$\,$}

\def\sles{\lower2pt\hbox{$\buildrel {\scriptstyle <}
   \over {\scriptstyle\sim}$}}

\def\sgreat{\lower2pt\hbox{$\buildrel {\scriptstyle >}
   \over {\scriptstyle\sim}$}}
\def\sharpnull#1{}

\null\voffset=-1.0pc  

\begin{document}
\def \m {$ M_\odot$~}
\def \l {$ L_\odot$}
\def \ro {g/cm$^{3}$}
\def \etal {\it et~al. \rm}
\def \f {{\nu}}
\def \pos {{\mathbf r}}
\def \vel {{\mathbf v}}
\def \dir {{\mathbf \Omega}}
\def \dpsidt {{1 \over c} {\partial \Psi \over {\partial t}}}
\def \dphidt {{1 \over c} {\partial \phi \over {\partial t}}}
\def \dpsidr {\mu {\partial \Psi \over {\partial r}}}
\def \dpsidx {{d \Psi \over {dx}}}
\def \dpsidz {\eta {\partial \Psi \over {\partial z}}}
\def \dpsidfi {\xi {\partial \Psi \over {r \partial \phi}}}
\def \dpsidmu {{(1-\mu^2) \over r} {\partial \Psi \over {\partial \mu}}}
\def \divr  {{1 \over r^2} {{\partial(r^2 \mu \Psi)} \over {\partial r}} }
\def \divx {{1 \over r} {{\partial(r \mu \Psi)} \over {\partial r}} }
\def \divmu {{1 \over r} {\partial((1-\mu^2) \Psi) \over {\partial \mu}} }
\def \divfi {{1 \over r} {\partial(\xi\Psi) \over {\partial \phi}}}
\def \dt    {\Delta t}
\def \cdt   {{1 \over {c \Delta t}}} 
\def \dtau  {d\tau}

\slugcomment{\bf}
\slugcomment{Submitted to Ap.J.}

\title{On the Sensitivity of Deflagrations in Chandrasekhar Mass White Dwarf to Initial Conditions}

\author{Eli Livne\altaffilmark{1},  Shimon M. Asida\altaffilmark{1},
 Peter H\"oflich\altaffilmark{2}}

\altaffiltext{1}{Racah Institute of Physics, The Hebrew University, Jerusalem,  Israel}
\altaffiltext{2}{Department of Astronomy, The University of Texas, Austin, UT \ 85721}

\begin{abstract}

We analyze the sensitivity of the flame propagation in a Chandrasekhar mass white dwarf
to initial conditions during the subsonic burning phase (deflagration), using 2D simulations
of the full WD.  Results are presented for a wide variety of initial flame distributions including
central and off-center single point and multi-point, simultaneous and non-simultaneous, ignitions.
We also examine the effects of convective velocity field which should
exist at the core before the thermo-nuclear runaway.

 Our main conclusion suggests that the amounts of burning products and their distributions through
 the deflagration phase are extremely
sensitive to initial conditions, much more sensitive than presented in previous studies.
In particular, we find that more complex configurations such as even slight off-center ignitions, 
non-simultaneous multi-point ignitions and velocity fields tend to favor solutions in which 
individual plumes rise faster than the bulk of a typical Rayleigh-Taylor driven, unstable
burning front. The difference to previous calculations for an octant of a WD may be understood
as a consequence of the suppression of l=1,2 modes. Our results are consistent with full star
calculations by the Chicago group.
Moreover, the total amount of nuclear burning during the phase of subsonic burning depends sensitively on
the initial conditions and may cause the WD to pulsate or to become unbound.
We discuss the implications of the results on current models for Type Ia SNe,
limitations imposed by the 2-D nature of our study, and suggest directions 
for further study.

\end{abstract}


\section{Introduction}
\label{intro}

 The last decade has witnessed an explosive growth of high-quality
data for thermonuclear explosions of a White Dwarf Star (WD), the Type Ia Supernovae (SNe~Ia).
 Advances in computational methods provide new insights into the physics of the
phenomenon and  a direct, quantitative link between observables and explosion physics.
 Both trends combined provided spectacular results, allowed to address, to identify 
specific problems and to narrow down  the range of scenarios.
 However, with the advances came the realization that observational constrains seem to be
at odds with the most elaborated calculations for deflagration fronts, one of the central parts
in the currently most favored model, the thermonuclear explosion a Chandrasekhar mass WD.

 The $M_{Ch}$ scenario requires an initial phase of  pre-expansion because,
otherwise, almost the entire WD would burn to $^{56}Ni$, in contradiction to the 
observation which show a significant amount of intermediate mass elements, namely O, Si, and S.
  This pre-expansion is commonly believed to occur during an initial phase of a slow
deflagration that preserves the  WD structure but decreases the binding energy. 
In absence of strong mixing, the pre-expansion depends mainly on the total amount of 
burning but  hardly on its actual rate \cite{dominguez00}.
 Successful spherical models need either a rapidly increasing deflagration speed and no
radial mixing (similar to  W7  \cite{nomoto84} -- see footnote
\footnote{The pure deflagration model W7 is a spherical model and, consequently, 
shows a layered structure which is typical for detonations. Realistic  3-D 
deflagration models do not show a radial layering of the abundances.}),  or a
deflagration-detonation transition (DDT).
 For 'successful' models, $\approx $ 0.2-0.3 $M_\odot$ are burned during the deflagration phase leading to a 
loosely but still bound WD ($E_{WD,pot} \approx  1-2 \times 10^{50} erg$)

The hydrodynamical evolution of a WD undergoing burning by subsonic narrow flame is governed by Rayleigh-Taylor 
(RT) instability and the consequent buoyancy of hot plumes of ashes. This buoyant motion has two major effects -
first, it increases the front surface and therefore enhances the effective burning rate. At the same time the
spherical distribution of different elements is being destroyed and a complex structure, 3D by nature, is formed. 
Therefore, multidimensional simulations became essential for establishing a viable reactive hydrodynamical model
 for the deflagration phase.
Recently, significant progress has been made toward the understanding of the flame physics of deflagrations.
 Starting from static WD, hydrodynamic calculations  of the deflagration fronts have been performed in 2-D 
(\cite{liv93l,n95,rein99,lisewski00})  and 3-D  (.e.g. \cite{khokhlov95,k01,A2,g02,G3}).
 As a common result, RT instabilities govern the morphology of the burning front in the regime of linear
 instabilities,
i.e.\ as long as  perturbations remain small.  An initially spherical flame propagates outward with a laminar
 velocity of about 100 km/sec, becomes
distorted by Rayleigh-Taylor instabilities, and forms multiple plumes at different scales. Hot, burned matter
plumes rise outward due to buoyancy, and cold, unburned matter is pulled toward the inner region.
Nuclear burning is ongoing and, as a result, the plumes mainly consists of iron-group elements with a skin
of intermediate mass elements produced when the plume reaches low density regions. Overall, the effective burning speed
is typically less than  10 \% of the sound speed.  The energy release by plumes will result in the expansion of the 
entire envelope and, eventually, the Ni-plumes will 'freeze' out in the expanding envelope. 
Outer layers of the envelope expand close to the speed of sound and cannot be reached by the burning front.
 In pure 3D-deflagration models, typically 1/2 to 2/3 of the total mass undergoes nuclear burning and releases
about 1. to 1.3 $10^{51} erg$ of nuclear energy, and the WD becomes unbound, and produces a 'healthy explosion'.
 All multi-dimensional, pure deflagration SN models seem to show the same properties, namely a thick layer of unburned
C/O rich matter, a non-radially stratified structure of the burning products,  and explosion energies
about half of that is required to fit spectra and LCs. 

The three dimensional nature is starting to become accessible by direct observations of
remnants (Fesen 2004), spectro-polarimetry (\cite{howell01,wang03}), early time IR-spectra for
 subluminous SNe~Ia, and  late-time spectral line profiles (e.g.\cite{b97,hetal03}).
 Some 3-D effects seem to be unavoidable for a system originating
from  a close binary system with mass overflow because of its interaction with the accretion disk
and the companion star (\cite{liv92,marietta00}). Other possible reasons include instabilities in the burning front,
namely during the deflagration phase, off-center detonations or the transitions from deflagration to detonation, 
and the rapid rotation of the progenitor. We start to see evidence for all these asymmetries with the noticeable
exception for the Ni plumes or, more precisely, we do not see these signatures using the various methods,
although they should have been seen. For recent reviews see \cite{branch99,h05}.

 Assuming the runaway develops to a deflagration front, the sensitivity of the burning history to initial
 conditions was first revealed by \cite{nim1}, where a set of several
 initial ignition configurations have been used for 2D simulations. However, the set was very limited
and left open many questions.
In general, the authors found that the buildup time of RT-instabilities is shorter corresponding to the
larger gravitational acceleration,  producing an off-center distributions of iron-group elements.
 Still the basic characteristics of deflagration burning remain unaltered, namely the  chemical distribution
and the typical amount of burning.
 However, one case was unique. In case of an ignition in a single point at 200 km off-center, 
an isolated hot bubble rises very quickly to the surface, leaving a bound star.
 Recently, based on 3D calculations,  Calder et al. (2003) found similar behavior when ignition occurs only 12 km
 off-center in an spherical bubble of 50 km in radius.
  We shell refer to this phenomena as \it{ the single bubble catastrophe }\rm,
 as in this case the deflagration ends with a dud explosion. In such a case, only a few percents
 of the star are burned and a large plume of ashes reaches very early the surface of the WD.
 
 Unfortunately, we do not know what should be proper initial conditions for the deflagration phase.
 The evolution of the WD toward thermonuclear runaway and the initial position of the
 flame are hardly known. H\"oflich and Stein (2002) were the first to perform
 2D simulations of the last minutes before the runaway. Those simulations show that ignition occurs at
 a single point off center, at a radius of 30-50 km. They found that generations of rising plumes
 create a convective velocity field which, by and large, dominates the initial motion of the
individual plumes. As a consequence, the  transition from non-explosive to explosive burning, the ignition,
 occurs most likely when a hot plume is dragged downward and compressed. Based on 3-D calculations 
prior to the runaway, Woosley \& Wunch (2004) suggested that large scale velocity pattern is 
being established on time scales of several minutes which causes hot plumes to be carried outward and,
as a consequence, may favor larger off-center ignitions. It is under discussion, whether this global pattern survives the
rapid increase of nuclear burning during the last minutes prior to runaway (\cite{h05}).

 In this work, we want to study the range of parameters that span the set of very different combustions,
between the \it{ single bubble catastrophe }\rm, and the partially 'healthy' explosion. We therefore
investigate the influence of a wide variety of initial conditions using our best
numerical tools in a parametric manner. We focus on the properties of the deflagration phase and the observational
consequences.  We vary the number of ignition points and their
 distribution in space and time. Furthermore, in part of the simulations we take into account 
 a pre-existing velocity field which should develop during the convective burning stage which
 precedes the runaway. In particular, a major subject of this study is to understand the conditions at
 which single bubbles are formed and escape to the surface.
  Despite the limitations of 2D simulations to give quantitative results to the problem at
 hand, we believe that qualitatively our 2D results are valid, and should be used in future
 similar 3D studies.

\section{Basic Methods and Tools}
\label{basic}

We have used the 2D code VULCAN (\cite{liv93}), with new capabilities listed below,
to simulate the deflagration phase in an isentropic Chandrasekhar mass WD. 
The code is based on cylindrical coordinates and is used for many years to simulate
astrophysical flows under the assumption of axial symmetry.
Since this is a comparative study which focus on the effects of initial conditions, 
the details of other numerical and physical components of the simulations are not
important, provided the simulations are being performed close enough to the 'realistic'
case. For example, the initial composition of the WD has important impact on the final
abundances, but for our purposes it has only minor effect on the flow and we impose
in all the presented simulations the same C-O ratio, namely 1. The isentropic structure
of our initial configurations is a very good approximation to a more realistic structure
due to two reasons. Firstly, the entropy has little effect on the pressure in the very
degenerate part (close to the center) of the star. Secondly, a large part of the star
is convective at the final stages of evolution so that an almost isentropic 'convective core'
develops around the center. The equation of
state is the standard 'eos' for degenerate electrons + radiation + nuclei gas.
Simplified reactions network with approximate binding energy have been used (see
below more details). 

We use 'nearly' polar grids (but cylindrical coordinates) with varying radial spacing and constant
angular spacing. The grid in most of the simulations consists of 150 radial zones
(logarithmically spaced) and 150 angular zones. Verification runs with 100 and 200 zones
in each direction gave very similar results in which the larger patterns are the same and
the total energy production vary by a few percents (see figure 5d). In order to avoid the singularity of
the polar grid we generate a special finite element grid inside an inner radius of 100 km,
 which is connected smoothly to the main polar grid. Contrary to conventional polar grids, our hybrid
grid enables us to follow transversal motions through the center with little numerical expenses.
The only price we pay here is that the spurious numerical errors of the hydrostatic equilibrium
are a bit larger in this inner zone. We have tested hydrostatic
equilibrium by running the code without burning and found no significant expansion or contraction
after a whole one second. Some spurious velocities are seen in the inner zone, where the grid is not
polar, but even there those velocities are much smaller than velocities generated by the deflagration.

Several sets
of initial conditions, ignition points, were used in order to study the wide range
of the unknown ignition parameters. The first set consists of ignition in a sphere
of a given radius. This case is similar to the initial conditions used in \cite{G3}.
A second set is based on igniting a single point off center on the axis, at a given
distance from the center ( see \cite{chicago}). In a third set we ignite 30 points
distributed in a given volume in time and space. Here we extend similar but simultaneous
initial conditions used in \cite{ropke04} to non-simultaneous multi-point cases. The
temporal distribution of the points in the non-simultaneous case is exponential with a
time scale of 0.1 sec, as suggested in \cite{W1}. 

Our ignition method does not introduce any abrupt disturbance to the star (which
could happen due to careless initial energy deposition at finite ignition regions).
In fact, we start always from zero-volume ignition points or point, from which the
burning fronts propagate a short distance with some prescribed initial speed. This
speed is fast enough to carry the front away from the ignition points in a reasonable
time but slow enough to allow the star to react to the release of energy. The distance
on which this initial speed is used is very small, of the order of several grid cells.
Only in the first set of simulations, where we want to ignite a sphere of finite radius,
we increase this initial radius to the radius of the initial sphere. 	
From there the fronts propagate according to the basic front algorithm described below. 

For each set of ignition points we computed two simulations. In the first one we start
from a stationary configuration in hydrostatic equilibrium. In a second simulation we
added initially a convective velocity field which mimics the early convection which
precedes the runaway. The magnitude of this field is comparable to the laminar speed
near the center (see subsection 2.2). 
The simulations presented here are limited to a finite time due to grid limitations.
Our grids follow the global expansion of the star but we can not follow at the
moment the high speed rise of small bubbles after they reach the surface of the star.
To follow the evolution further an extended grid is required, like in \cite{chicago}.
Nevertheless, the conclusions are valid for the early phase deflagrations and delayed
detonation models.

\subsection{Front Propagation and Burning}
\label{front}

A special algorithm for front propagation have been developed. Using a special interface
tracking algorithm, we utilize the multi-phase capability of the code to improve the
accuracy of our simulations. This algorithm minimizes numerical dissipation in the advection
step and at the same time enables to define and to advance the location of the front in a
very accurate way. We use a 'Volume of Fluid' (VOF) method (see review in \cite{vof}) to
locate the interface between fuel
and ashes, at any given time, in cells containing two phases. Given the partial volumes of
those two phases in a cell, we construct the interface in such a way that it is perpendicular to
the gradient of phases and divide the cell into two parts with the given partial volumes.
This algorithm, which is very accurate, is used twice in each time step - in the advection
(remapping) step and in the burning step, where the propagation of the front is calculated.

The front propagates perpendicular to itself at a speed which is the maximum between the
laminar speed and a turbulent speed based on a sub-grid model. We adopt here the simple
RT driven formula for the turbulent speed $v_t=0.5 \sqrt{Agl}$, where $A$ is Atood's
number, $g$ is the gravitational acceleration and $l$ is a length comparable to the local
mesh size. This recipe was first used in this context by Livne (\cite{liv93l}) and later was
calibrated  by A. Khokhlov (\cite{khokhlov95}) using 3D control simulations. More complicated
recipes for turbulent speed, which include Kelvin-Helmholtz instability effects at shear
layers (\cite{n95}), are now under investigation, but that study lies beyond the scope of this work.
We may comment that uncertainties in the turbulent speed may change the energy production by at least
$10^{50}$ ergs, while they only weakly affect how the ashes are distributed in the ejecta. Those
uncertainties remain because both numerical calibrations (\cite{khokhlov95}) and terrestrial experminets 
(\cite{n95}) were not performed in spherically expanding envelopes. 

The binding energy is being released in the volume swept by the front locally.
Because we are not interested here in detailed nucleosynthesis, and in order to
enable the performance of many simulations, the simulations presented here were calculated
with simplified rates, which produce accurate enough 'q-value' (the amount of
energy released per gram) at various densities.
A control simulation where an alpha network have been used gave similar results.  
Both the alpha network and the simplified rates take into account photo-dissociation,
which is the important back-reaction at high density. Thus, the 'q-value' increases from
roughly $3 \times 10^{17} erg/g$ at the center to about $6\times 10^{17} erg/g$ at the low
 density regions.
This q-value underestimates the alpha network q-value at low densities by a few percents,
so that the buoyancy in the simulations is certainly not overestimated.

\subsection{Early Convection}
\label{convection}

It was shown numerically (\cite{yossi}) and theoretically that a convective velocity field
evolves before the runaway, with convective speed comparable to the laminar speed of the
deflagration in its first stages. Consequently, convective speed of 100 km/sec
 (with typical Mach number of 0.02 at the center) is expected at runaway. However,
2D simulations of the early convection are not only very expensive but they also do not
 produce a correct hierarchy of scales. To overcome this, we construct
a 2D artificial initial velocity field, imposed on the flow at t=0. The construction uses
multi-mode potential flow which is incompressible in the sens that $div(\rho \bf{v})=0 $.
 The field is generated by imposing a vorticity source which has zero total circulation
 and is distributed in a given region of the star (inside the convective core).
 With no 3D simulations of the pre-runaway
phase, we can only guess the amplitudes and wavelengths of the source
according to the qualitative 2D fields obtained by H\"oflich \& Stein 2002. 
We found that indeed those fields do not introduce artificial pressure waves and mimic
nicely initial (yet unknown) incompressible convective field (see fig.1). 
Further details will be reported elsewhere. 

\placefigure{fig1}

\section{2D Simulations - Results}
\label{results}

 A few snap shots from simulations with a single point ignition are shown in figure 2.
 Regardless the location of the ignition point they all end up with a single bubble
 which reaches the surface at a speed of nearly 5000 km/sec. The bubble accelerates mainly
 after about 0.5 second, when it rises into the region having large pressure and gravitational
 slopes ( small pressure scale height). In those cases the deflagration incinerates
 only a few percents of the fuel and the WD remains bound. The effect of early convection
 in this case is very small and do not change the evolution. Larger distance between the
 center and the ignition point leads to smaller amount of burning. In large, our results
 are in very good agreement with the results shown in \cite{chicago}.   

\placefigure{fig2}

 The simulations which start with a burning sphere of radius 30 km show different results
 (figure 3A). Although they all develop bubbles and spikes under the RT instability no
 single bubble gain enough speed to overcome the global expansion. The nearly homologous
 expansion keeps the burning zone confined to the central region and burning proceeds at
 relatively high density. As a result, the amount of energy released is enough to unbind
 the star. The fact that the obtained energy in our simulation is too small for healthy
 SNe~Ia can be attributed to the well known difference between 2D and 3D simulations. In
 this sense our results agree with previous 2D and 3D simulations (see introduction). When initial convection
 is being introduced into the simulation the outcomes change by small amount energy-wise but a very
 large asymmetry of bubbles develop after one second (figure 3B). This finding should have
 important implications concerning observables of such models (see conclusions). 

\placefigure{fig3}

 Multi points ignition cases yield intermediate results, in between the two extreme cases
 described above. In all those cases we find large asymmetries and effective burning rates which are
 significantly smaller than those of the spherical case. The asymmetries, which go together with
 reduced burning rates, are larger when non-simultaneous ignition is employed or/and when initial
 convective velocity field is introduced at t=0. It is enough that one of these physical components
 will be present in a simulation to induce large asymmetry at late time. Figure 4 show some snap
 shots of the bubbles distribution after about 1.2 seconds for multi-point cases. One can see that
 in some cases a single bubble does float to the surface while at other cases the bubbles remain 
 confined to the inner part of the expanding envelop. The main parameter which distinguish between the
 two cases is the location of the ignition point which runaway first.

\placefigure{fig4}

Figures 5a-5d summarize the energy production rates for the simulations 
discussed above. In figure 5a we present the results of the simulation with
a single off-center ignition. We can see that the energy production is low and only small 
differences exist between the different cases. The only exception is the
case in which the ignition is relatively close to the center (at a radius of 
10 kilometers), where more energy is produced.
In the simulations with initial burning sphere, energy production exceeds the
binding energy, as can be seen in figure 5b. The existence of initial
convective velocity field does not affect significantly the energy production
(at least until the time in which the first bubble reaches the surface).
In figure 5c we can see that in the case of multi-point ignition, initial
convective velocity field reduces the energy production. A non-simultaneous
ignition has a greater effect as well. Figure 5d presents the sensitivity of
the energy production graph to the resolution of the grid for the simulation 
with simultaneous multi-point ignition (compare to figure 4A).

\placefigure{fig5}

\section{Conclusions}
\label{conclusion}

The outcomes of a deflagration process in a Chandrasekhar mass WD are determined 
primarily by the distribution in space and time of the initial flamelets. Different
distributions lead to very different evolutions, from dud events to 'reasonable'
explosions. The two extreme cases are the runaway of a single point on one side and
the spherical volume ignition on the other side. Both cases are not very likely to
happen in nature because of the complicated convective structure of the flow before
runaway. The single point case leads always to a dud event in which only a few percents
of the star are consumed. Whether or not a \it confined detonation \rm (\cite{chicago})
follows such an event should be examined in full 3D simulations. In this case however it
is assumed that no other point runaway during the first second which follows the ignition
of the first point. On the other side of the spectrum spherical volume ignition leads
to significant burning which unbind the star. In 2D, the energy released is too small to
produce acceptable SN but 3D simulations were shown to release much more energy, on the lower
side of healthy explosions. The presence of initial convective velocity field show little
effect in this case on the total energy released but induce a large asymmetry on the burning
region. We may argue that the volumes initially ignited in published 3D simulations are
much too big to be realistic. When the radius of the initial burning sphere is reduced to
a few kilometers the behavior of the following deflagration becomes more similar to the case
of a single point.

In the case of multi-point ignition two effects play important roles. First,
the amount of energy released and the structure of the bubbles are very sensitive to the spatial
and temporal distribution of the local runaways. With simultaneous multi-point ignition the
amount of energy released drops by twenty percent compared to volume ignition and the star is
marginally bound (limited to 2D). When non simultaneous multi-point ignition is used we get
another drop of twenty percent in the binding energy released, which makes the explosion model
a dud event. The second effect is the presence of initial convective velocity field, which causes
a similar drop in the total burning and similarly harm the explosion. We notice however that those two
effects are not additive and their joint effect is similar to the effect of each of them. The reason
for the drop in energy release in those cases is that once a single or several bubbles escape with high speed
the total effective burning rate drops significantly, because burning on the surface of those bubble occurs
at very low density. 

At large, we find a very diverse space of possible distributions of burning products, as function of initial
conditions. Large asymmetries and large variations in burning energy and products are found. 
Non-simultaneous multi-point ignitions and velocity fields tend to favor solutions in which
individual plumes (dud solutions) rise rather than being dominated the typical Rayleigh-Taylor driven, unstable
burning fronts. In case of dud-solutions, our results agree well with those of \cite{calder03}.
 However, in other 3D simulations, the single bubble catastrophe was not observed.
We can suggest qualitative explanation to those differences.
Gamezo \etal  (\cite{G3}) start the simulation with a spherical hot bubble of 30 km radius, which
maintains hydrostatic equilibrium. However, since they compute only one octant with reflecting boundary
conditions the most important low unstable modes are suppressed. R\"opke \& Hillebrandt (2004) start their
simulation with a large number of simultaneous ignition points, distributed inside a radius of 200 km. This
causes prompt expansion which tends to suppress the buoyancy of single plumes. Furthermore, their spatial
resolution of $\delta x = $ 8 km may be too coarse to allow the instability to develop. In our 2D simulations
the grid size at $ r=$ 100 km , which is the radius where the plumes start accelerating, is roughly 2 km.  

 It is beyond the scope of this work to provide
a general overview over the relation between observables, possible DDT and the influence of variations 
in the progenitor. For this purpose, we would like to refer to recent reviews and articles 
(\cite{branch99,hetal03,G3,h05}). Instead, we want to put  specifically our results in the context
 of explosion models, and to evaluate possible observational consequences. 
We show that deflagration phases vary widely in their energy production from 'dud' events with
 nuclear energy productions from $10^{50}erg$ to healthy explosions, and everything in between.
This opens up the possibility that the WD undergoes a phase of one or multiple pulsations, a solution
which previously has been excluded (\cite{A2}). Furthermore, the possibility to trigger a detonation by
the Zeldovich mechanism during pulsations (\cite{k92,hkw95}) can no longer be ignored.  
 As mentioned in the introduction, spherical explosion models require a pre-expansion of the WD which
requires nuclear burning between $0.2-0.3M_\odot$ before a DDT (or a phase of fast deflagrations) occurs.
Based on this, it has been suggested (\cite{hetal03,h05})
 that the DDT should occur during the expansion phase, which is
RT dominated. Based on temperature fluctuations scales it has been argued that 'self-induced' DDT
 is unlikely in this phase (\cite{niemeyer99}), and some other ('external') mechanisms have been suggested such as
 shear flow induced instabilities in the layers between initial WD and accreted matter
(\cite{hetal03,yoon04a,yoon04b}). However, the results shown here should now re-open
that discussion by considering non-radial pulsational modes and large scale circulation that
may have time to develop and trigger a detonation.

In summary, the results of this work show larger sensitivity of the deflagration products to initial
conditions at runaway than assumed or reported before. Therefore, it is urgent to verify these results
in a full star 3D simulations (which are unfortunately beyond the available computer capabilities of the
authors). Even if only partially verified (smaller sensitivity in 3D ?) they cast serious questions
concerning the currently standard mechanism for SNIa's and call for more work focused on pre-runaway
evolution. 

\acknowledgments

We would like to acknowledge discussions with all our colleges. E. Livne thanks P. Mazzali,
J. Stein and A. Glasner for fruitful discussions. 
This work has been supported by the NASA grants (NAG5-7937, HST-GO-10118.05-A)
and by NSF grant AST0307312 to PAH.


\clearpage

\figcaption[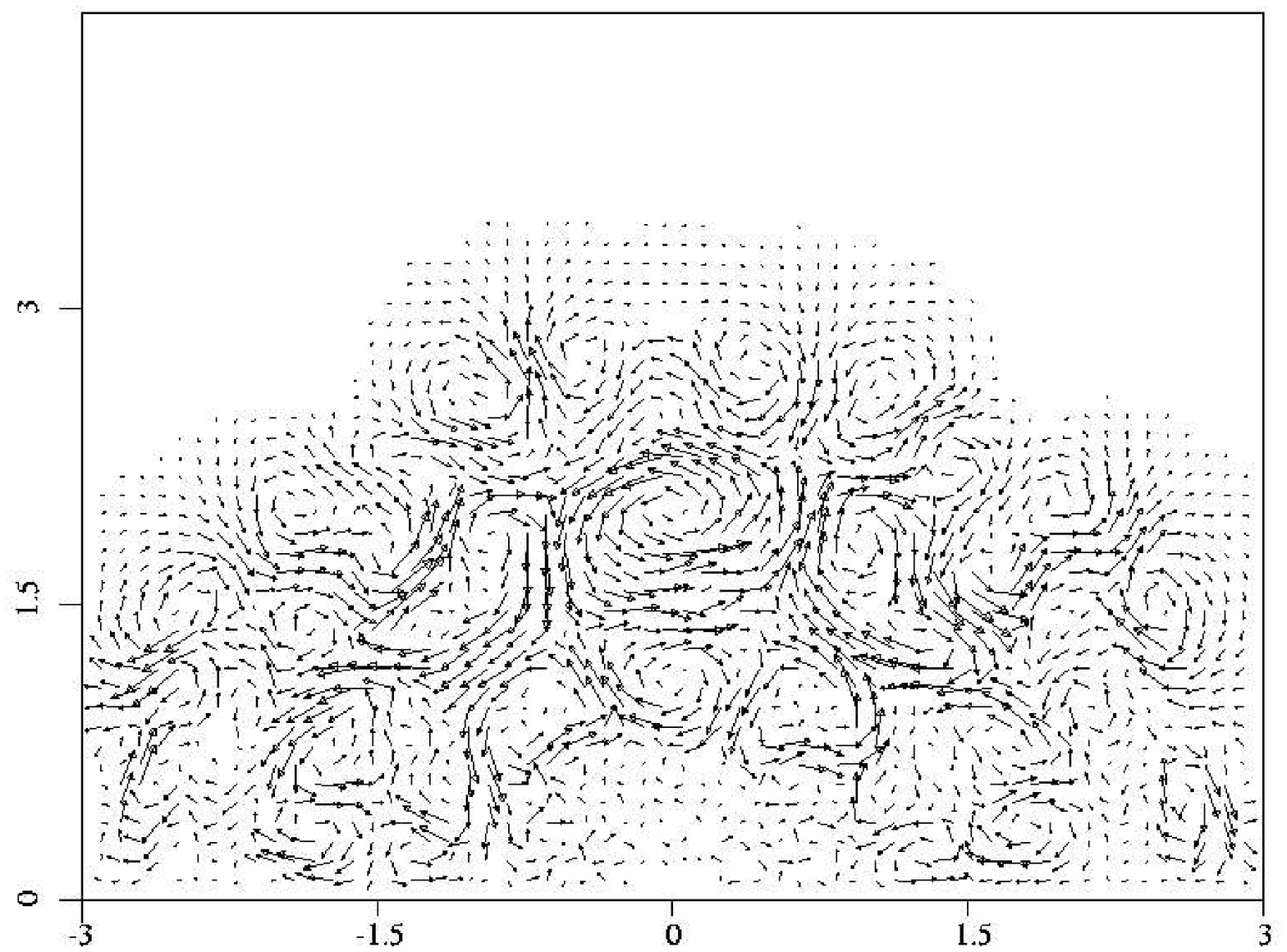]{Initial convective velocity field. The longest arrows correspond
to a velocity of 160 km/s. Only central region of star is presented. Unit scale
on axes is 100 km. \label{fig1}}

\figcaption[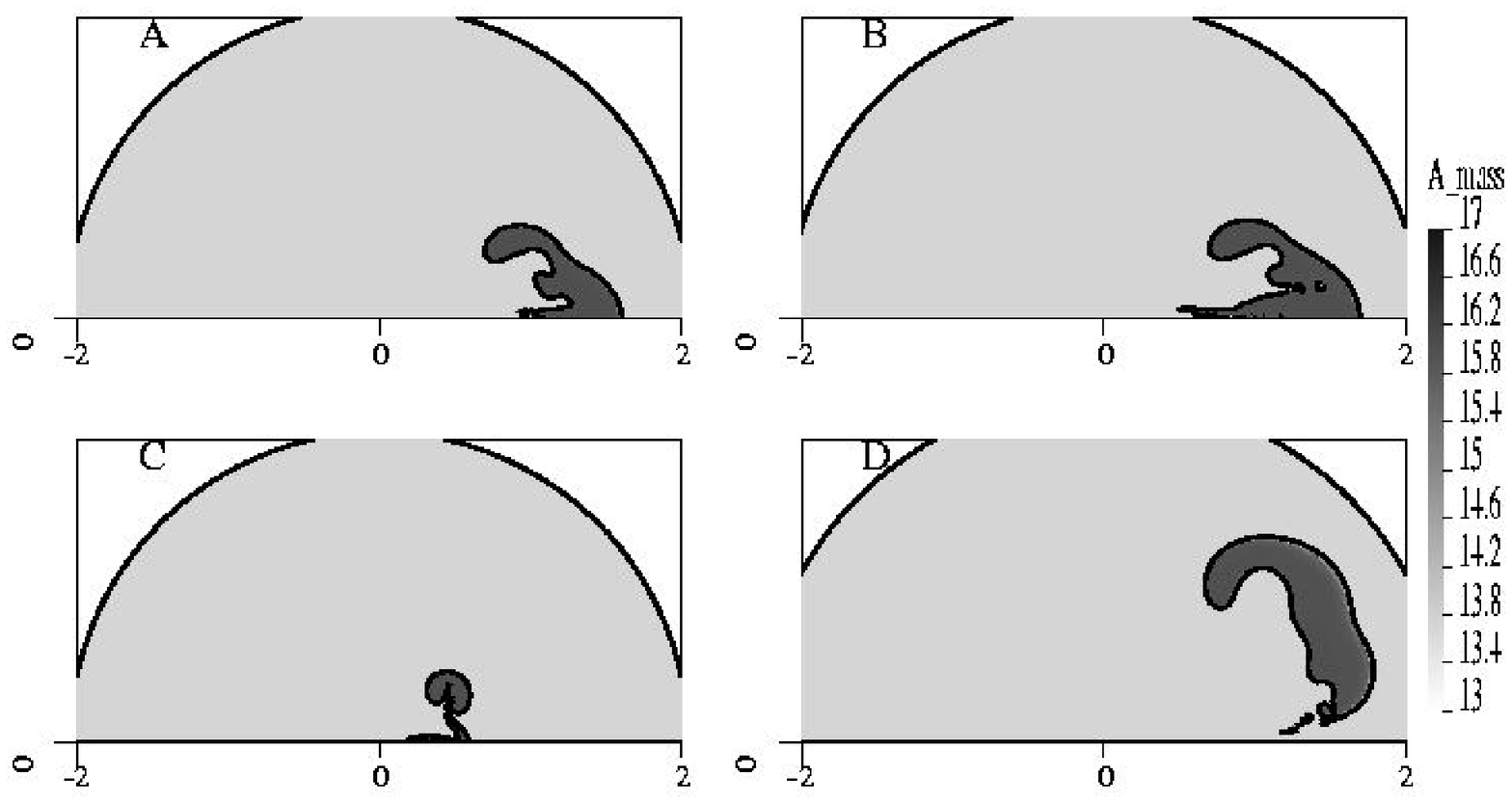]{Burning region of simulations with single off-center ignition, at
T=1.2 s. Unit scale on axes is 1000 km. Four simulations are presented:
A - Ignition at r=20 km . 
B - Ignition at r=20 km with initial convective velocity field.
C - Ignition at r=10 km . 
D - Ignition at r=100 km . \label{fig2}}

\figcaption[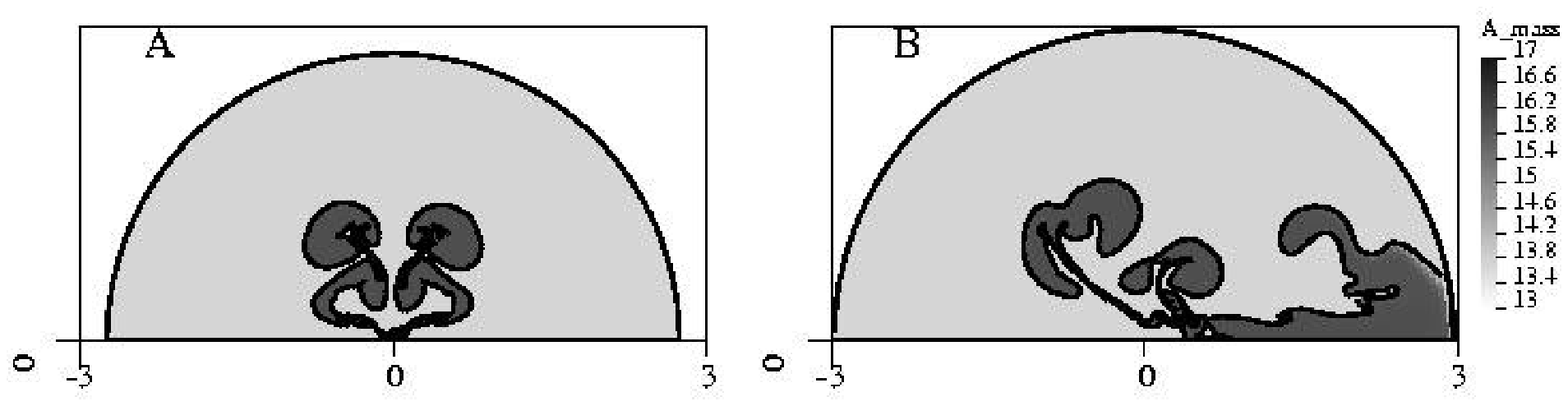]{Burning region of simulations with central  sphere ignition, at
T=1.5 s. Unit scale on axes is 1000 km. Two simulations are presented:
A - no initial convective velocity field.
B - with initial convective velocity field. \label{fig3}}

\figcaption[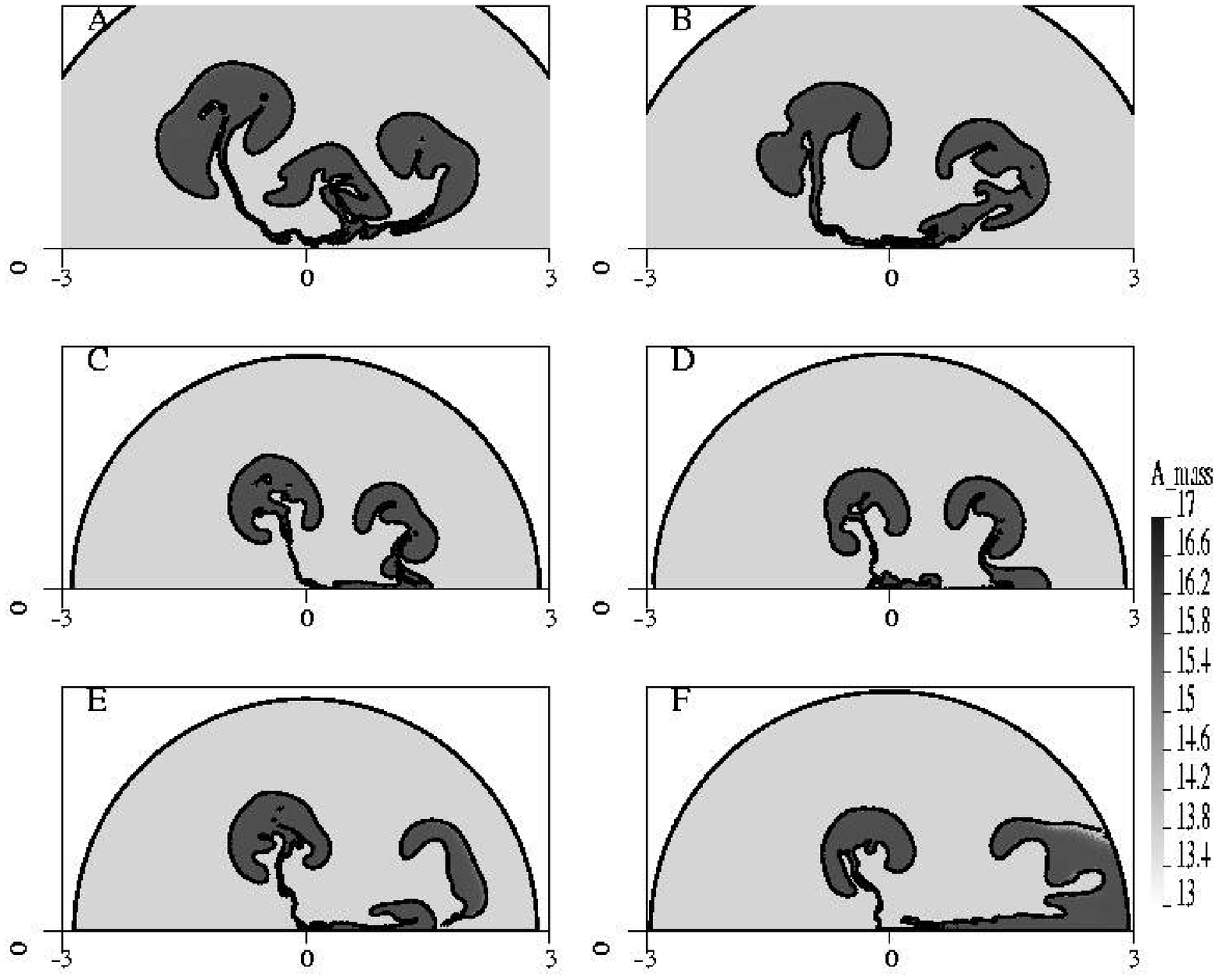]{Burning region of simulations with multi-point ignition, at
T=1.2 s. Unit scale on axes is 1000 km. six simulations are presented:
A - simultaneous ignition.
B - simultaneous ignition with initial convective velocity field.
C - non simultaneous ignition.
D - non simultaneous ignition with initial convective velocity field.
E - non simultaneous ignition, location of first ignition point on axis.
F - non simultaneous ignition with initial convective velocity field, 
    location of first ignition point on axis. \label{fig4}}

\figcaption[fig5.eps]{Total energy vs. time plot. Each simulation is marked by the figure
number in which it's burning region is presented above.
a - comparing single off-center ignition simulations.
b - comparing central  sphere ignition simulations.
c - comparing multi-point ignition simulations.
d - Energy production in the simultaneous multi-point simulation (fig. 4A)
    with three different resolutions. \label{fig5}}

\clearpage

\end{document}